\documentclass{article}
\usepackage[dvips]{graphicx}

\begin{document}
\title{A Multifractal Description of Wind Speed Records}
\author{Rajesh G. Kavasseri \\ Department of Electrical and Computer
Engineering \\ North Dakota State University, Fargo, ND 58105 -
5285 \\ ~(email: rajesh.kavasseri@ndsu.nodak.edu) \\   \\
Radhakrishnan Nagarajan \\ University of Arkansas for Medical
Sciences, Little Rock, AR 72205}
\date{}

\maketitle

\begin{abstract}
\noindent In this paper, a systematic analysis of hourly wind speed
data obtained from four potential wind generation sites in North
Dakota is conducted. The power spectra of the data exhibited a power
law decay characteristic of $1/f^{\alpha}$ processes with possible
long range correlations. The temporal scaling properties of the
records were studied using  multifractal detrended fluctuation
analysis {\em MFDFA}. It is seen that the records at all four
locations exhibit similar scaling behavior which is also reflected
in the multifractal spectrum determined under the assumption of a
binomial multiplicative cascade model.
\end{abstract}

\section{Introduction}
Wind energy is a ubiquitous resource and is a promising
alternative to meet the increased demand for energy in recent
years. Unlike traditional power plants, wind generated power is
subject to fluctuations due to the intermittent nature of wind.
The irregular waxing and waning of wind can led to significant
mechanical stress on the gear boxes and result in substantial
voltage swings at the terminals, \cite{fariley}. Therefore, it is
important to build suitable mathematical techniques to understand
the temporal behavior and dynamics of wind speed for the purposes
of modeling, prediction, simulation and design. Attempts to
identify the features of wind speed time series data were
described in \cite{haslett79} and \cite{raftery82}. To our
knowledge, the first paper to bring out an important feature of
wind speed time series was \cite{raferty_memory}. In
\cite{raferty_memory}, the authors examined long term records of
hourly wind speeds in Ireland and pointed out that they exhibited
what is known as long term dependence. Seasonal effects, spatial
correlations and temporal dependencies were incorporated to build
suitable estimators. Evidence for the presence of long memory
correlations was provided by inspecting the periodogram of the
residuals from a fitted and autoregressive model of order nine
i.e. AR(9)
\cite{raferty_memory}. \\

\begin{figure}[htbp] \centering
\includegraphics[height=4in, width=4in]{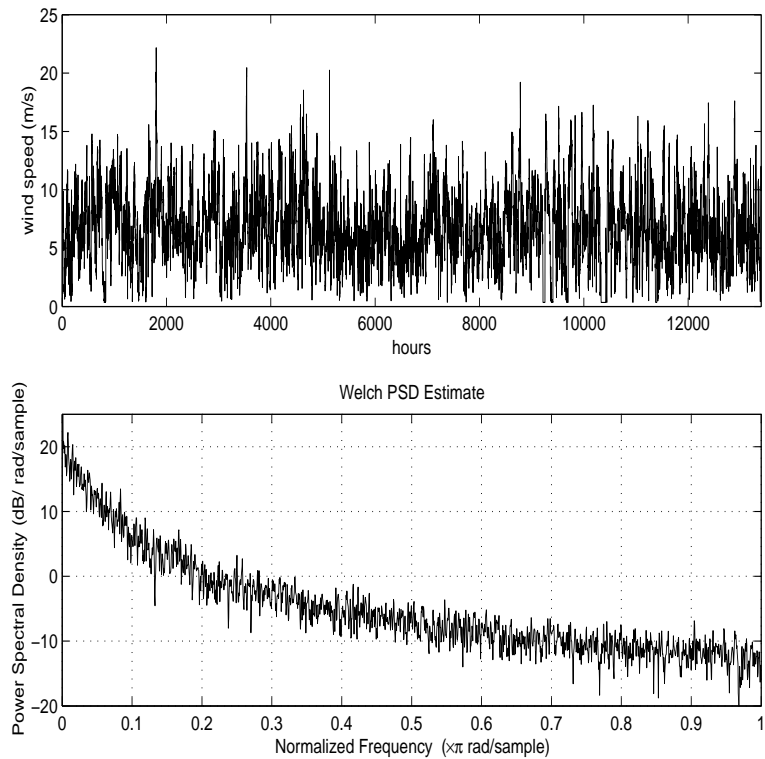}
\caption{Temporal trace and Power spectrum of a representative
record} \label{trace_psd}
\end{figure}

\noindent For the pursuit of wind energy development in the state
of North Dakota which has an estimated potential of 250 GW,
\cite{randall}, several wind monitoring stations have been set up
where the hourly average wind speeds along with meteorological
variables are being recorded. As the power developed by the wind
turbines is proportional to the cube of the wind speed, the
recorded data can be used to assess the wind power potential of
the site. In addition, an analysis of the recorded data can be
useful in understanding the nature of the fluctuations in power
developed by the wind turbine. Conventionally, the statistical
modeling of wind speed data has been done by using Weibull and
Rayleigh probability distribution functions, \cite{shetty}. While
these methods are useful in providing estimates of wind generated
energy yield, they do not explicitly bring out the nature of the
temporal variations in wind speed and thus, wind power. Complex
motions in the atmosphere tend to render the wind speed distinctly
non-stationary and intermittent. Traditional power spectral (or
equivalently the auto correlation function) analyses are
restricted in their scope for the present study due to their
susceptibility to non-stationarity. Power spectral techniques have
been useful successfully to detect possible long-range
correlations of the form of the form $S(f) \sim 1/f^{\beta}$,
\cite{feder88}. Long range correlations generally indicate that
temporally well separated samples of the time series are
correlated with each other and indicative of self-similar
(self-affine) behavior. Self similar time series can be
characterized by

\begin{equation}
y(t) \equiv a^{\alpha}y(t/\alpha) \label{defn_selfsimilar}
\end{equation}
where the $\equiv$ in Eqn.(\ref{defn_selfsimilar}) denotes that
both sides of the equation have identical statistical properties.
The exponent $\alpha$ in Eqn.(\ref{defn_selfsimilar}) is called
the self-similarity parameter, or the scaling exponent.  Values of
$\alpha$ in the range (0, 0.5) characterize anti-persistence,
whereas those in the range (0.5,1) characterize persistent long
range correlations with $\alpha = 0.5$ representing uncorrelated
noise. The exponent ($\beta$) estimated from the power spectrum
$(S(f) \sim 1/f^{\beta}$) is related to $\alpha$ as $\beta =
2\alpha -1 $. The temporal trace and corresponding power spectrum
of one of the representative records is shown in Fig.
\ref{trace_psd}.

\noindent Examining the power spectrum of the wind speed indicates
a power-law decay of the form $S(f) \sim 1/f^{\beta}$. The
auto-correlation functions (ACF) decay slowly to zero and the
first zero crossing of the ACFs occur at lags of 61,56,60 and 57
respectively for the four data sets. Such features are
characteristic of statistically self similar processes with well
defined power law correlations, \cite{feder88}. \\

\noindent  Estimating the Hurst exponent ($H$) from the given data
is an alternate effective way to determine the nature of the
correlations in it \cite{hurst}. Hurst exponents have been
successfully used to quantify long range correlations in plasma
turbulence \cite{plasmaturb1}, \cite{plasmaturb2}, finance
\cite{hurstfinance1}, \cite{hurstfinance2}, network traffic,
\cite{selfsimilar_traffic} and physiology, \cite{ivanov_nature}.
Long range correlations are said to exist if $0.5 < H < 1$ (see
\cite{beran}, \cite{bassing} for details). There are several
methods such as the variance method, re-scaled range (R/S)
analysis, Whittle estimator and wavelet based methods to estimate
the Hurst exponent. Hurst estimators are susceptible to such
artifacts such as polynomial trends which cannot be ruled out in
experimental data and therefore, they may give spurious results.
Recently the Detrended Fluctuation Analysis (DFA), \cite{peng1994}
and its extensions have been proposed as an alternate effective
technique to determine possible long range correlations in data
sets obtained from a diverse settings,
\cite{peng1995}, \cite{hausdorff1995}, \cite{ausloos97}, \cite{ivanova_1}. \\

\noindent It should be noted that the techniques listed above can
only extract a single scaling exponent from a time series.
However, it is possible that the given process may be governed by
more than one scaling exponents, in which case a single scaling
exponent would be unable to capture the complex dynamics inherent
in the data. Therefore, these methods are appropriate only for the
analysis of monofractal signals which have uniform scaling
properties throughout the signal which can be characterized by a
single exponent. On the other hand, multifractal signals are  far
more complex than monofractal signals and require more than one
(theoretically infinite) exponent to characterize their scaling
properties,
\cite{ivanov_nature}. \\

\noindent In the present study, the wind speed time series data is
studied using a fairly robust and powerful technique called {\em
Multifractal Detrended Fluctuation Analysis} (MFDFA),
\cite{kantel_mfdfa}. The method provides a systematic means to
identify and more importantly quantify the multiple scaling
exponents in the data \cite{kantel_mfdfa}. The scaling exponents
of the data are estimated under the assumption of a binomial
multiplicative cascade model. This is carried out on the wind
speed data acquired from four spatially separated monitoring
locations in North Dakota. The paper is organized as follows. In
Sec.\ref{dacq}, the acquisition of wind speed data is described.
In Sec. \ref{mfdfa}, a brief description of multifractal detrended
fluctuation analysis is provided. In Sec.\ref{results}, the
results of {\em MFDFA} and the calculation of the multifractal
spectrum is provided. The conclusions are summarized in
Sec.\ref{conclusions}.

\section{Data Acquisition}
\label{dacq} 
\begin{figure}[htbp]
\centering
\includegraphics[width=2.5in]{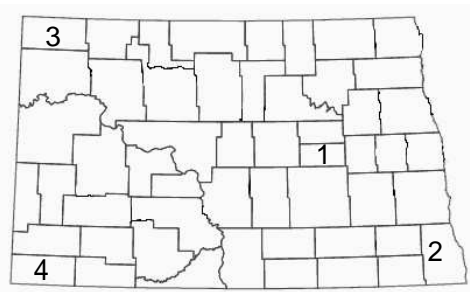}
\caption{A map of the wind monitoring sites} \label{map_ND}
\end{figure}

The hourly averaged wind speeds at four
geographically well separated locations in North Dakota were
recorded by means of a conventional cup type anemometers located
at a height of 20 m. The co-ordinates of the map are provided in
Table \ref{site_coords}. From Fig.\ref{map_ND}, it can be seen
that the locations are sptially well separated across North
Dakota. The wind speed data to be discussed were recorded over a
time frame ranging from 11/29/2001 to 07/28/2003.

\begin{table}[htbp]
%% increase table row spacing, adjust to taste
\renewcommand{\arraystretch}{1.0}
\caption{Co-ordinates of the wind monitoring sites}
\label{site_coords} \centering
%% Some packages, such as MDW tools, offer better commands for making tables
%% than the plain LaTeX2e tabular which is used here.
\begin{tabular}{|c|c||c|c|}
\hline Station & Latitude & Longitude & Elvn (ft)  \\
\hline Site 1 & N 47 27.84' & W 99 8.18' & 1570
\\ \hline Site 2 & N 46 13.03' & W 97 15.10' & 1070 \\
 \hline Site 3 & N 48 52.75 & W 103 28.4' & 2270 \\
\hline Site 4 & N 46 12.67' & W 103 13.07' & 2880 \\ \hline
\end{tabular}
\end{table}

\section{Multifractal Detrended Fluctuation Analysis (MFDFA)}
\label{mfdfa} MFDFA, \cite{kantel_mfdfa} a generalization of DFA
has been shown to reliably extract more than one scaling exponent
from a time series. A brief description of the algorithm is
provided in Section III, for completeness. A detailed explanation
can be found elsewhere \cite{kantel_mfdfa}. Consider a time series
$\{x_k\}, k = 1 \dots N$. The MFDFA algorithm consists of the
following steps.
\begin{enumerate}
\item The series $\{x_k\}$ is integrated to form the integrated
series (also called ``profile") $\{y_k\}$ given by
\begin{equation}
y(k) = \sum_{i=1}^{i=k} [x(i) - \bar{x}] \;\;\; k = 1, \dots N
\end{equation}
where $\bar{x}$ represents the average value.

\item The series $\{y_k\}$ is divided in to $n_s$ non-overlapping
boxes of equal length$s$ where $n_s = int(N/s)$. To accommodate
the fact that some of the data points may be left out, the
procedure is repeated from the other end of the data set
\cite{kantel_mfdfa}.

\item The local polynomial trend $y_v$ with order $v$ is fit to the data in each box,
the corresponding variance is given by

\begin{equation} F^2(v,s) = \{ \frac{1}{s} \sum_{i=1}^{i=s}
\{y [N-(v-n_s)s + i] - y_v(i)\}^2
\end{equation}

for $v = 1, \dots n_s$. Polynomial detrending of order m is
capable of eliminating trends up to order m-1. \cite{kantel_dfa}

\item The $qth$ order fluctuation function is calculated from
averaging over all segments.

\begin{equation}
F_q(s) = \left \{ \frac{1}{2 n_s} \sum_{i=1}^{i= 2
n_s}[F^2(v,s)]^{q/2}  \right \} ^{1/q}
\end{equation}
In general, the index $q$ can take any real value except zero.

\item Step 3 is repeated over various time
scales $s$. The scaling of the fluctuation functions $F_q(s)$
versus the time scale $s$ is revealed by the log-log plot.

\item The scaling behavior of the fluctuation functions is
determined by analyzing the log-log plots $F_q(s)$ versus $s$ for
{\em each} $q$. If the original series $\{x_k \}$ is power-law
correlated, the fluctuation function will vary as
\begin{equation}
F_q(s) \sim s^{h(q)} \label{hq}
\end{equation}
\end{enumerate}

\subsection{Multifractal Analysis}
\noindent In the MFDFA procedure (Section III), the exponent
$h(q)$ describes the scaling behavior of the $qth$ order
fluctuation function. For positive values of $q$, $h(q)$ describes
the scaling behavior of segments with large fluctuations while
those of negative values of $q$, describe scaling behavior of
segments with small fluctuations, \cite{kantel_mfdfa}. For
stationary time series, the exponent $h(2)$ is identical to the
Hurst exponent. Thus the exponent $h(q)$ is called as the
generalized Hurst exponent, \cite{kantel_mfdfa}. For monofractal
time series which are characterized by a single exponent over all
scales, $h(q)$ is independent of $q$, whereas for a  multifractal
time series, $h(q)$ varies with $q$. This dependence is considered
to be
a characteristic property of multifractal processes, \cite{kantel_mfdfa}.\\

\noindent A multifractal description can also be obtained from
considering partition functions ( see \cite{barabasi} for details)
\begin{equation}
Z_q(s) = \sum_{v=1}^{v=n_s} | y_{vs} - y_{(v-1)s} |^q \sim
s^{\tau(q)} \end{equation} where $\tau(q)$ is the Renyi exponent.
A linear scaling of  $\tau(q)$ with $q$ is characteristic of a
monofractal data, whereas a nonlinear scaling is indicative of
multifractal behavior. The $h(q)$ obtained from MFDFA is related
to the the Renyi exponent $\tau(q)$  by

\begin{equation}
q h(q) = \tau(q) + 1 \label{tauq}
\end{equation}

\noindent Therefore, another way to characterize a multifractal
series is the multifractal spectrum $f(\alpha_{h})$ defined by,
\cite{feder88}
\begin{equation}
\alpha_{h} = \frac{d \tau(q)}{d q},\;\; f(\alpha_{h}) = q
\alpha_{h} - \tau(q)
\end{equation}
In (9), $\alpha_{h}$ is the H\"{o}lder exponent which
characterizes the singularities (cusps, ridges, chirps, spikes) in
a process $X(t)$ at time $t$, \cite{reidi_asilo}. The multifractal
spectrum $f(\alpha_{h})$ describes the singularity content of the
process, i.e. the dimension of the set of times $t$ where
$\alpha_h = \alpha$, \cite{reidi_asilo}. Substituting for
$\tau(q)$ from Eqn.(\ref{tauq}), we get, \cite{kantel_mfdfa}

\begin{equation}
\alpha_{h} = h(q) + q \frac{d h(q)}{dq}, \;\; f(\alpha_{h}) =
q[\alpha_{h} - h(q)] + 1 \label{alphaq}
\end{equation}

\noindent The generalized exponents $h(q)$ in Eqn.(\ref{alphaq})
can be estimated by the formula
\begin{equation}
h(q) = \frac{1}{q} - \frac{ln(a^q + b^q)}{q ln2} \label{mfcascade}
\end{equation}
assuming a binomial multiplicative cascade model \cite{feder88}
which has served as one of the standard paradigms to describe
multifractal scaling, (see \cite{kantel_river}, \cite{feder88},
\cite{kantel_mfdfa} for details). Finally, the width of the
multifractal spectrum $f(\alpha_{h})$ at $f = 0$, given by
\begin{equation} \Delta \alpha_{h} =
h(-\infty) - h(\infty) = \frac{(ln(b) - ln(a))}{ln2}
\label{mfwidth}
\end{equation}
is computed to compare the strength of multifractality for
different records. Incidentally, the multifractal spectrum can
also be estimated using wavelets by the wavelet transform modulus
maxima (WTMM) method, \cite{arneodo} which we do not pursue here.

\section{Results}
\label{results} 
\begin{figure}[htbp]
\centering
\includegraphics[height=4in, width=4in]{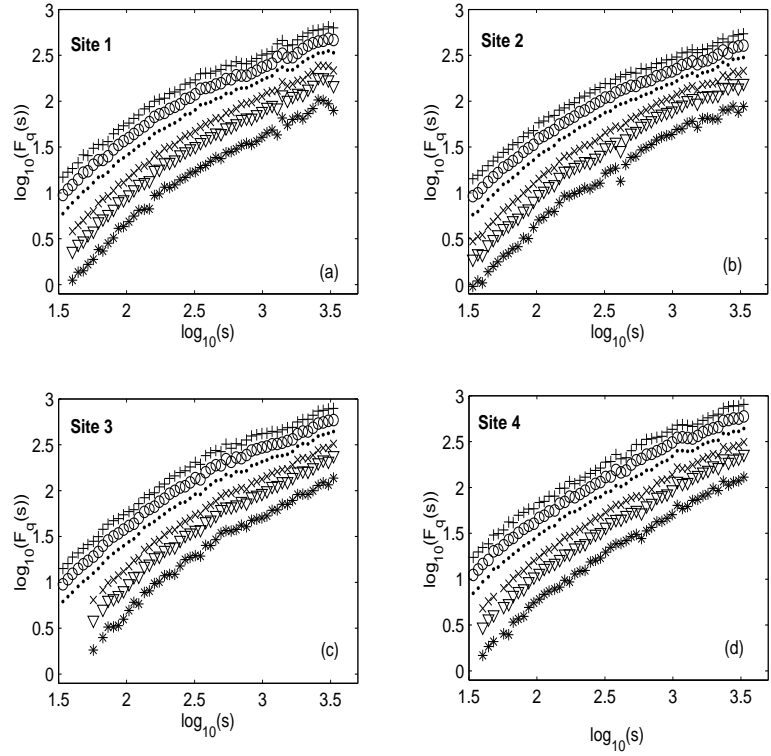}
\caption{Fluctuation functions at all sites using MFDFA. Symbols
used to indicate the various moments are $q = -6$ (-),
$q=-4$(inverted triangle), $q =-2$ (x), $q = 2$ (.), $q = 4$ (o), $q
= 6$ (+). The curves are vertically shifted for clarity.}
\label{mfdfa_allsites}
\end{figure}

\begin{figure}[htbp] \centering
\includegraphics[height=3in,width=4in]{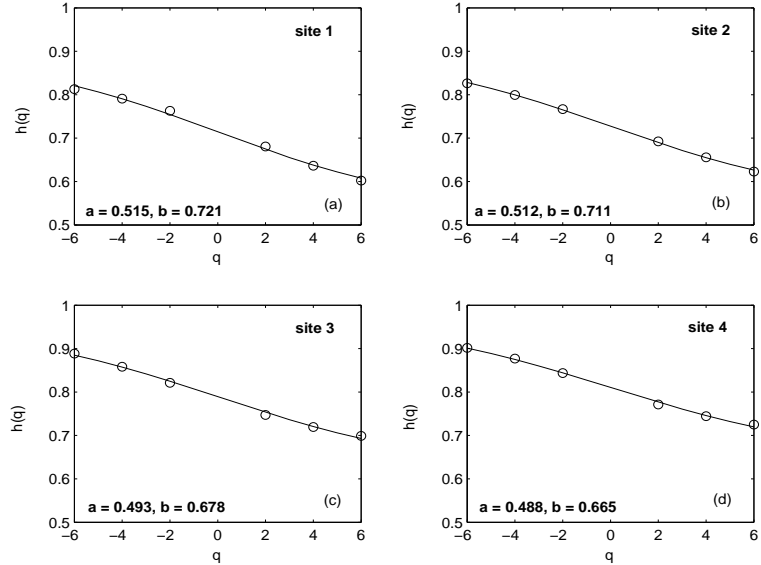}
\caption{Generalized scaling exponents for wind speed records at
the four sites. Circles indicate the actual values of the scaling
indices for $q = -6,-4,-2,2,4 \;$and$\;6$. The solid line
represents the fit from the  multiplicative cascade model
(Eqn.(\ref{mfcascade})) obtained from a nonlinear least squares
procedure. The values of $a$ and $b$ are indicated for each site.}
\label{hq4sites}
\end{figure}

\noindent The log-log plot of the fluctuation
functions $F_q(s)$ vs $s$ for the wind speed records at all four
locations with (q = -6, -4, -2, 2, 4, 6) using fourth order
polynomial detrending is shown in Fig.\ref{mfdfa_allsites}.  The
generalized hurst exponent $h(q)$ estimated for (q = -6, -4, -2, 2,
4, 6) using Eqn.(\ref{hq}) is shown in Fig.\ref{hq4sites}. One can
note that at all locations, the slopes $h(q)$ decrease as the moment
is increased from negative to positive values. For example, the
slope decreases from 0.88 when $q = -6$ to 0.6989 when $q = +6$ for
site 3 which is indicative of multifractal behavior.

\begin{figure}[htbp] \centering
\includegraphics[height =4in, width=4in]{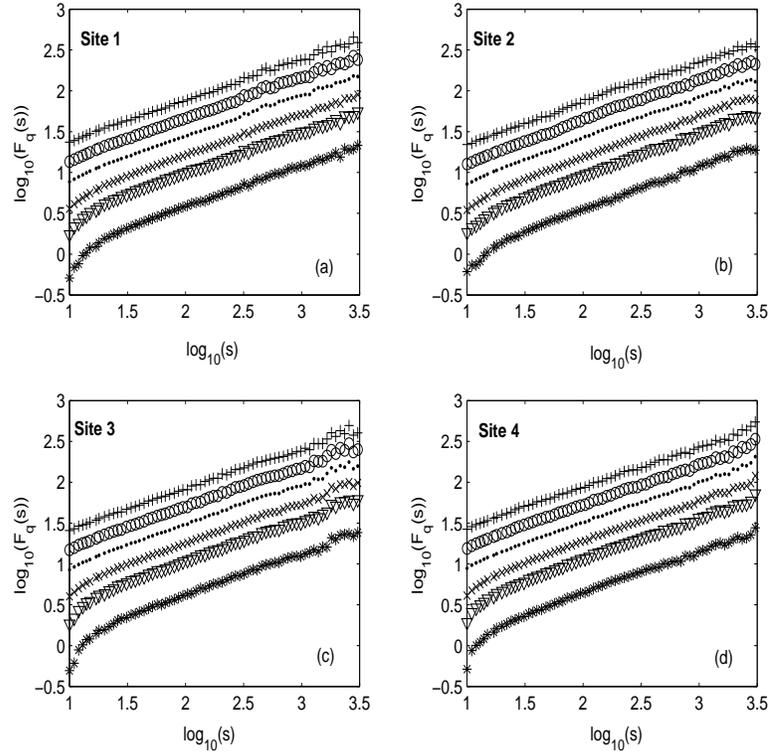}
\caption{Fluctuation functions at all sites using MFDFA for random
shuffle surrogates of the original data. Symbols used to indicate
the various moments are $q = -6$ (-), $q=-4$(inverted triangle),
$q =-2$ (x), $q = 2$ (.), $q = 4$ (o), $q = 6$ (+). The curves are
vertically shifted for clarity. Note that the curves are parallel
to each other with a slope of $\sim 0.5$. } \label{rshuff_mfdfa}
\end{figure}

\begin{figure}[htbp] \centering
\includegraphics[height=3in,width=3in]{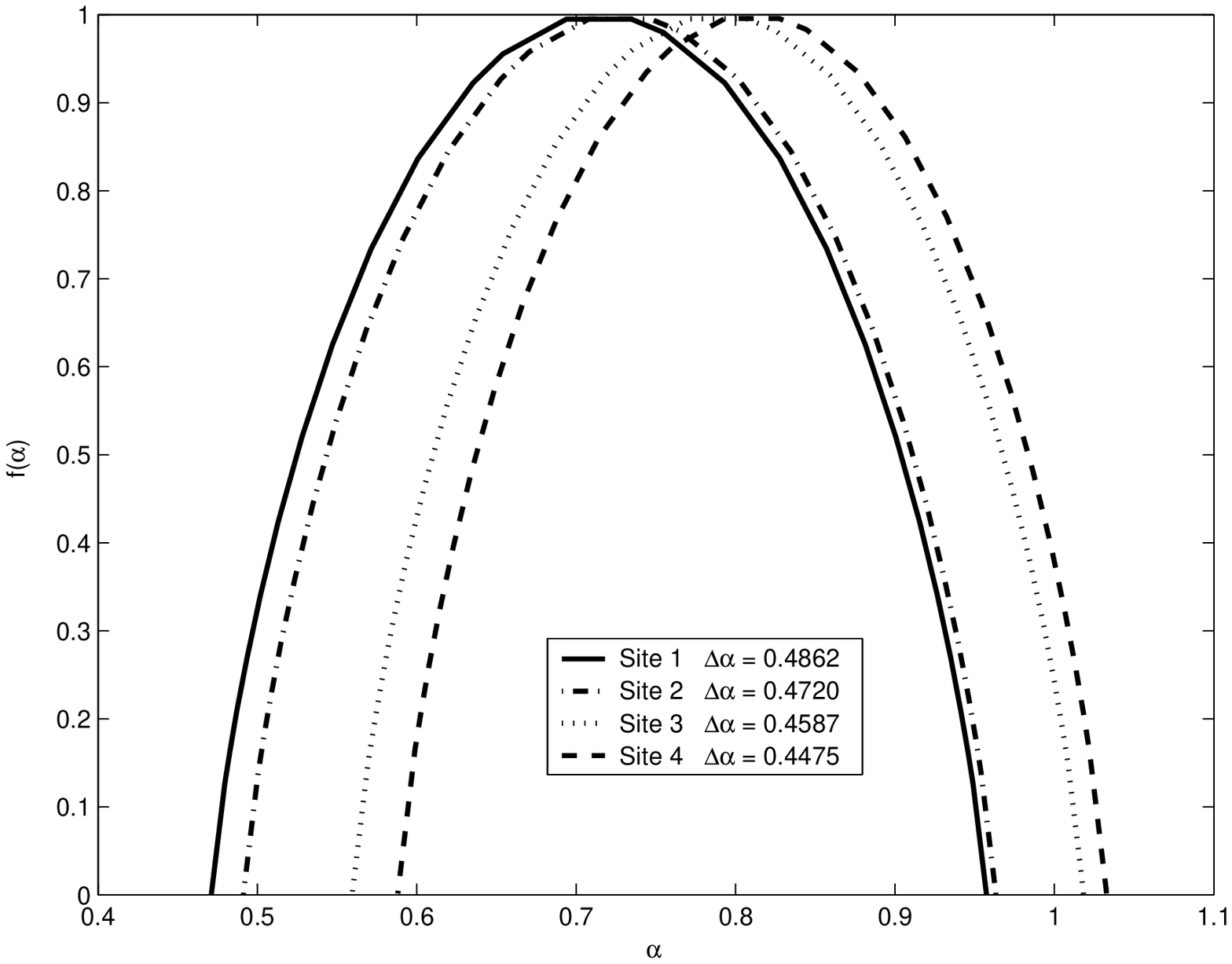}
\caption{Multifractal spectra of wind speed records at the four
sites } \label{multispec}
\end{figure}

\noindent Recent studies have suggested the choice of constrained
randomized shuffled surrogates to reject the null that the observed
scaling is due to the distribution as opposed to the temporal
structure in the data. Constrained randomized shuffles (surrogates)
are similar to traditional bootstrapping without replacement, i.e.
each member in the original data is retained in the surrogate. While
the temporal structure is destroyed the distribution of the original
data is retained in the surrogate realizations. The surrogate
realizations are equivalent to uncorrelated noise. A good exposition
of the concepts of surrogate analysis can be found in \cite{small}.
In the present study, the MFDFA studies are performed on random
shuffle surrogates of the original data, to reject the claim that
the observed scaling exponent obtained on the original data is due
to its distribution and not due to correlations present in it. The
results shown in Fig. \ref{rshuff_mfdfa} indicate that for all
values of $q$, the fluctuations scale as $F_q(s) \sim s^{0.5}$ which
signifies a loss of correlations. In other words, there is a loss of
multifractality because the slopes $h(q)$ are independent of $q$.
The parameters $a$ and $b$ determined under the assumption of a
binomial multiplicative process Eqn.(\ref{mfcascade}) is also shown
Fig.\ref{hq4sites}. Using the values of $a$ and $b$ with
Eqns.(\ref{hq}), (\ref{alphaq}) and (\ref{mfcascade}), the
multifractal spectrum at all four locations shown in
Fig.\ref{multispec} is computed. The width of the multifractal
spectrum is calculated from Eqn.(\ref{mfwidth}).

\section{Conclusions}
\label{conclusions} In this study, the hourly wind speed records at
four geographically well separated locations in North Dakota were
examined. The focus was on determining the nature of the
correlations and hence, the dynamics in the wind speed data. The
power-law decay of the power spectrum suggested the presence of
possible long range correlations and self-similar behavior. A number
of techniques proposed in the past to quantify self-similarity in a
given data have been susceptible to non-stationarity which can
manifest itself as polynomial trends. This motivated us to use a
fairly robust technique such as the MFDFA. Analysis using MFDFA
elucidated the dependence of $h(q)$ on $q$, which is a hall mark of
multifractal processes. The binomial multiplicative cascade model
has been widely used to determine the multifractal spectra. The
multifractal spectra $(f(\alpha))$ and the widths $(\Delta \alpha)$
(assuming a binomial multiplicative cascade) were estimated at the
four
locations.\\

\noindent Temporal and spatial variations of wind speed are
influenced by several factors such as pressure gradient, turbulence,
temperature and topography. Our preliminary studies suggest that
long term wind speed variations possess a multifractal structure.
Moreover, despite the  heterogeneities across the spatially
separated locations, certain quantitative features such as the
spectrum ($f(\alpha)$) and multifractal widths $(\Delta\alpha$) seem
to be retained. A deeper investigation of these issues with
extensive data sets over wider geographical regions may provide
clues towards understanding the nature of long term atmospheric wind
speed variations.

\section*{Acknowledgment}
\noindent The financial support from ND EPSCOR through NSF grant
EPS 0132289 and services of the North Dakota Department of
Commerce Division of Community Services are gratefully
acknowledged.

%\begin{biography}[{\includegraphics[width=1in,height=1.25in,clip,keepaspectratio]{mshell}}]{Michael Shell}
% where an .eps filename suffix will be assumed under latex, and a .pdf suffix
% will be assumed for pdflatex; or if you just want to reserve a space for
% a photo:

%\begin{biography}{Michael Shell}
%\end{biography}

%\begin{biographynophoto}{John Doe}
%\end{biographynophoto}

% insert where needed to balance the two columns on the last page
%\newpage

%\begin{biographynophoto}{Jane Doe}
%\end{biographynophoto}


%\vfill


\end{document}